\documentclass{article}
\usepackage{spconf,amsmath,graphicx,hyperref}
\usepackage{url}
\usepackage{booktabs}
\usepackage{array}
\usepackage{amssymb}  
\usepackage{textcomp} 
\usepackage{color}
\usepackage{listings}
\usepackage{xcolor}
\usepackage{multirow}

\title{SynthCloner: Synthesizer-style Audio Transfer via Factorized Codec with ADSR Envelope Control}

\name{Jeng-Yue Liu$^{1,2,3*}$, Ting-Chao Hsu$^{1*}$,
      Yen-Tung Yeh$^{1}$, Li Su$^{2}$, Yi-Hsuan Yang$^{1}$}
\address{$^{1}$ National Taiwan University \quad $^{2}$ Academia Sinica \quad $^{3}$ Carnegie Mellon University}

\begin{document}

\maketitle
\begingroup
\renewcommand\thefootnote{}\footnotetext{*$\;$Equal contribution.}

\addtocounter{footnote}{0}
\endgroup

\begin{abstract}

Electronic synthesizer sounds are controlled by parameter settings that yield complex timbral characteristics and ADSR envelopes, making synthesizer-style audio transfer particularly challenging. Recent approaches to timbre transfer often rely on spectral objectives or implicit style matching, offering limited control over envelope shaping. Moreover, public synthesizer datasets rarely provide diverse coverage of timbres and ADSR envelopes. To address these gaps, we present SynthCloner, a factorized codec model that disentangles audio into three attributes: ADSR envelope, timbre, and content. This separation enables expressive audio transfer with independent control over these attributes. Additionally, we introduce SynthCAT, a new synthesizer dataset with a task-specific rendering pipeline covering 250 timbres, 120 ADSR envelopes, and 100 MIDI sequences. Experiments show that SynthCloner outperforms baselines on both objective and subjective metrics, while enabling independent attribute control. The code, model checkpoint, and audio examples are available at \url{https://buffett0323.github.io/synthcloner/}. 

\end{abstract}
\section{Introduction}
\label{sec:intro}
\begin{figure*}[t]
  \centering
  \includegraphics[width=\textwidth]{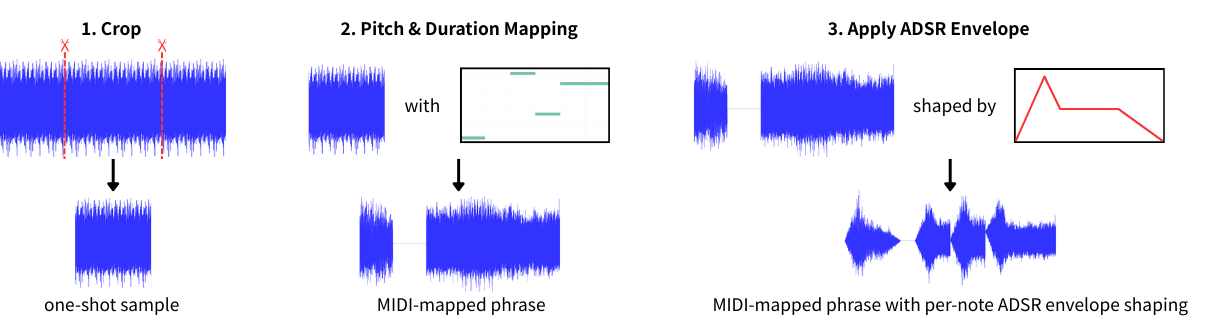}
  \vspace{-9mm}
  \caption{Data rendering pipeline of SynthCAT. A sustained-note segment is cropped, mapped to the MIDI sequence through pitch shifting and duration alignment, and then shaped with the ADSR envelope on a per-note basis to produce the final audio.}
  \label{fig:data}
\end{figure*}

Electronic synthesizers have become essential tools in modern music production, offering vast flexibility in sound design. The sound of a synthesizer is defined by a parameter configuration of oscillators, filters, and envelopes, that jointly shape its sonic characteristics. Despite the creative possibilities this offers, crafting settings remains labor-intensive and requires expert knowledge \cite{audiosynth_inversion}. 

To address this challenge, synthesizer-style audio transfer (SAT) aims to convert a source audio to match a reference while preserving the content, thereby reducing manual parameter tuning and enabling efficient sound exploration. In this paper, we define the sonic character of a basic synthesizer sound as comprising two aspects: timbre and ADSR envelope. Timbre denotes the \emph{time-invariant} spectral characteristics shaped by oscillators and filters; for simplicity, we exclude LFO-induced variations and additional envelope routing that introduce time-varying effects \cite{vail2014synthesizer}, while the ADSR envelope (Attack, Decay, Sustain, Release) governs the \emph{time-varying} master amplitude trajectory. Together, these components produce the sound that range from percussive plucks to sustained leads. Consequently, effective SAT requires accurate modeling and control of both timbre and ADSR envelope. 

Existing methods for timbre transfer \cite{latentdiffusionbridge, ssvae, musicti, cac, transplayer, ttimage, timbretron} focus on acoustic settings  or latent style mappings for reference-guided transfer, while guitar tone conversion \cite{zeroamplifier_guitar} models the analog signal chain to reshape tone. Both paradigms primarily emphasize spectral similarity while leaving the input’s ADSR envelope unchanged—an assumption valid for acoustic instruments but misaligned with synthesizer design. A related direction is synthesizer sound matching~\cite{audiosynth_inversion,synth_sound_matching,sound2synth,horner1993machine,yee2018automatic,2019inversynth,ssdiff_synth}, which recovers parameters through differentiable or DSP-based models; however, SAT requires direct learning from audio for timbre and envelope control. Recent work has begun to consider envelopes more explicitly: Caspe \emph{et al.}~\cite{fm_tone_transfer} predicted ADSR-like parameters to capture note onsets and offsets, while Masuda \emph{et al.}~\cite{ssdiff_synth} introduced a differentiable ADSR module. Yet these efforts focus on onset reproduction or differentiable control, rather than efficient SAT.

A further challenge is the lack of publicly available datasets that jointly span timbral and envelope diversity. NSynth~\cite{nsynth} offers large-scale note-level samples with broad timbre coverage but mostly sustained notes and limited envelope variation, while Synth1B1~\cite{onebillion} provides massive synthetic audio with parameter metadata but no explicit envelope control. To address this gap, we introduce SynthCAT, a dataset built from systematic combinations of 250 timbres, 120 ADSR envelopes, and 100 MIDI sequences, capturing the full diversity of combinations.

Building on this resource, we present SynthCloner, a factorized codec model inspired by FACodec~\cite{facodec}. SynthCloner disentangles synthesizer audio into three attributes, content, timbre, and ADSR envelope, and integrates information perturbation~\cite{NANSY} with supervised training~\cite{facodec} to enforce attribute separation. This enables faithful and controllable conversion across diverse sound, explicitly modeling both spectral and temporal dynamics.

The contributions of this paper are as follows:
\begin{itemize}
 
    \item We demonstrate empirically that explicit modeling of ADSR envelope control alongside timbre is essential for effective SAT.
    \item We present SynthCloner, the first model to our knowledge that performs SAT through factorization of timbre and ADSR envelope attributes.
    \item We introduce SynthCAT, a synthesizer dataset with 250 timbres, 120 ADSR envelopes, and 100 contents, providing a foundation for SAT.
\end{itemize}

\label{sec:data}

\vspace{-5mm}

\section{Data}
\vspace{-2mm}
To construct SynthCAT, we use the \texttt{Serum}\footnote{\url{https://xferrecords.com/products/serum-2}} synthesizer driven by MIDI notes, as Serum is one of the most widely used VST instruments in music production. Our goal is to render the same content under varied timbre and ADSR envelope settings. Isolating envelopes is non-trivial since many configurations couple them with oscillators, filters, or effects~\cite{lfo_dafx}, leading to unwanted timbral changes. To mitigate this, we select presets from commercial packages\footnote{\url{https://theproducerschool.com/}} that generate sustained tones and render them as long notes. From each rendering, we extract a one-second one-shot sample from the flattest region of the waveform, quantified by the flatness score~\cite{flatness,eff_vad}: 
\[
\text{Flatness}(x) = \frac{1}{1 + \mathrm{Var}(\mu(x))}, \quad 
\mu(x) = \tfrac{1}{N}\sum_{n=1}^{N} x[n],
\] 
where $x[n]$ is the waveform of length $N$, $\mu(x)$ is its temporal mean, and $\mathrm{Var}(\cdot)$ computes the variance across the segment. Segments with flatness above 0.95, indicating low variance in amplitude, are retained as distinct timbres. Each one-shot is then pitch-shifted and duration-aligned to MIDI sequences, and shaped with ADSR envelopes to render multi-note monophonic phrases. We generate 120 envelopes by uniformly sampling parameter ranges: Attack (10–100 ms), Decay (50–300 ms), Hold (0–200 ms), Sustain (0.0–0.80), and Release (30–300 ms). Finally, the dataset is built from the Cartesian product of 250 timbres, 120 envelopes, and 100 MIDI files from the mono-midi-transposition dataset~\cite{midi_data}, yielding 3M monophonic audio samples (44.1 kHz, mono) with a total duration of about 2,500 hours.

\begin{figure*}[!ht]
  \centering
  \includegraphics[width=\textwidth]{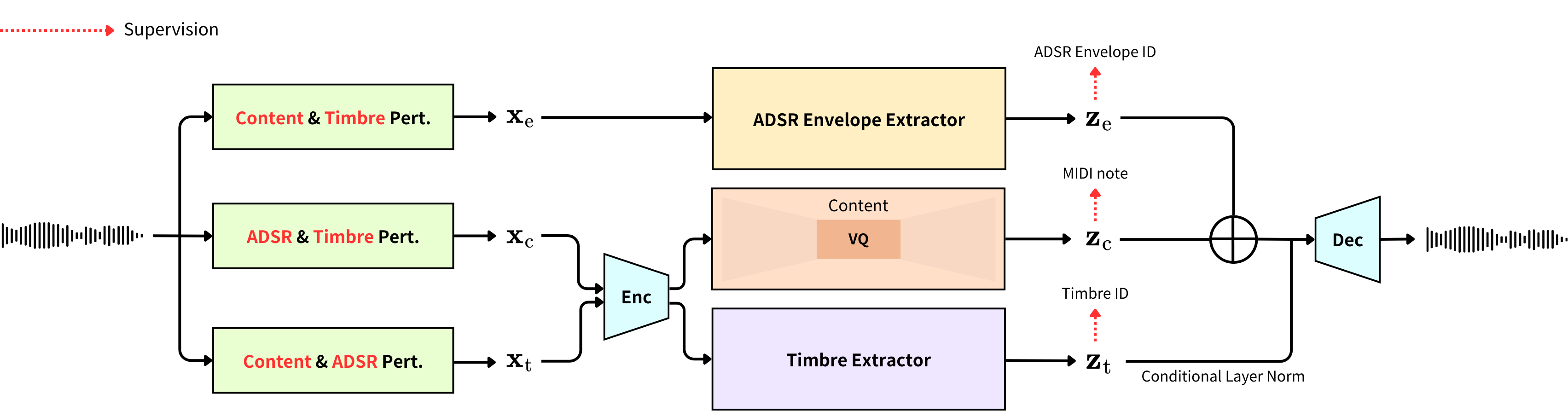}
  \vspace{-6mm}
  \caption{SynthCloner model architecture. To perform SAT, replace $\mathbf{x}_\text{e}$ and $\mathbf{x}_\text{t}$ with the reference audio.}
  \label{fig:model}
\end{figure*}

\section{Proposed Approach}
\label{sec:method}
\subsection{The SynthCloner Framework}
As illustrated in Fig.~\ref{fig:model}, SynthCloner is inspired by the FACodec \cite{facodec} framework but incorporates two key extensions: ADSR envelope modeling and attribute-specific information perturbation, detailed in Section \ref{ssec:disent}. The model decomposes an input audio $\mathbf{x}$ into three disentangled latent representations through separate processing paths, each with specific perturbations \cite{NANSY} during training.

First, the ADSR envelope path processes the perturbed input $\mathbf{x}_\text{e}$ by converting it to log-RMS value \cite{envelope_2011} and passing it through the ADSR envelope extractor which is a temporal multi-scale Conv-BiLSTM network to capture envelope dynamics \cite{conv_bilstm_vad}, yielding the ADSR envelope embeddings $\mathbf{z}_\text{e} \in \mathbb{R}^{D \times T}$, 
where $D$ denotes the embedding dimension and $T$ the frame length. Second, the content path follows the FACodec \cite{facodec} design, where the perturbed input $\mathbf{x}_\text{c}$ passes through the encoder and residual vector quantizer (RVQ) to produce the content embeddings $\mathbf{z}_\text{c} \in \mathbb{R}^{D \times T}$. Third, the timbre path, also adapted from FACodec \cite{facodec}, processes the perturbed input $\mathbf{x}_\text{t}$ through the shared-weight same encoder and a Conformer-based \cite{facodec, mfa_conformer} timbre extractor with global mean pooling, generating the timbre embedding $\mathbf{z}_\text{t} \in \mathbb{R}^{D}$ that captures global spectral characteristics.
During reconstruction, $\mathbf{z}_\text{e}$ is additively combined with  $\mathbf{z}_\text{c}$, then modulated by $\mathbf{z}_\text{t}$ through conditional layer normalization \cite{chen2021adaspeech}, before being decoded back to the waveform. For inference, the model enables SAT by using a reference audio as $\mathbf{x}_\text{e}$ and $\mathbf{x}_\text{t}$ to provide target timbre and ADSR envelope while preserving the original content.

\subsection{Attribute Disentanglement}
\label{ssec:disent}
While the three-path architecture provides a structural basis for attribute separation, it does not guarantee the disentanglement.
We introduce attribute-specific information perturbation, inspired by NANSY~\cite{NANSY}, to ensure that $\mathbf{z}_\text{e}$, $\mathbf{z}_\text{c}$, and $\mathbf{z}_\text{t}$ capture only their target attributes. During training, non-target attributes in the input are perturbed while the target attribute is preserved, forcing each encoder to produce embeddings invariant to irrelevant variations. SynthCAT, with its exhaustive combinations of timbre, ADSR envelope, and content, enables such perturbations. For example, when training the ADSR envelope encoder, given an input $\mathbf{x}$ with envelope $e_0$, content $c_0$, and timbre $t_0$, we construct a perturbed input $\mathbf{x}_\text{e}$ that shares $e_0$ but differs in content $c_1$ and timbre $t_1$, ensuring $\mathbf{z}_\text{e}$ remains invariant to non-envelope factors. Analogous perturbations are applied to the content and timbre paths. Following FACodec~\cite{facodec}, we further employ attribute-specific auxiliary tasks for disentanglement encouragement. 
For content embeddings $\mathbf{z}_\text{c}$, we use MIDI-derived pitch labels for frame-level supervision. For ADSR envelope and timbre embeddings ($\mathbf{z}_\text{e}$, $\mathbf{z}_\text{t}$), we apply categorical classification over 120 envelope IDs and 250 timbre IDs, respectively.

\begin{table*}[t]
\centering
\begin{tabular}{lcccccc}
\toprule
 & \multicolumn{3}{c}{Objective Metrics} & \multicolumn{3}{c}{Subjective Metrics} \\
\cmidrule(lr){2-4} \cmidrule(lr){5-7}
 & MSTFT$\downarrow$ & LRMSD$\downarrow$ &
   F0RMSE$\downarrow$ & TMOS$\uparrow$ &
   ADSRMOS$\uparrow$ & CMOS$\uparrow$ \\
\midrule
Ground Truth & -- & -- & -- & 4.08 & 3.96 & 4.25 \\
\midrule
SS-VAE~\cite{ssvae} & 7.22 & 0.92 & 641.62 & 2.20 & 2.25 & 3.41 \\
CTD~\cite{cac}      & 5.69 & 0.89 & 583.01 & 2.34 & 2.48 & 1.86 \\
\midrule
SynthCloner (ours) & \textbf{3.00} & \textbf{0.17} & ~~\textbf{20.64} &
\textbf{3.91} & \textbf{3.94} & \textbf{4.11} \\
-- w/o ADSR envelope path & 3.84 & 0.42 & ~~29.04 & 3.09 & 2.40 & 3.76 \\
\bottomrule
\end{tabular}
\caption{Objective and subjective results for synthesizer-style audio transfer }
\label{tab:main_results}
\end{table*}

\label{ssec:result}
\section{Experiment Setup}

\subsection{Training Setup} 
\noindent \textbf{Dataset}.
For evaluation, we hold out 250 seen timbres along with 20 unseen envelopes and 10 unseen MIDI sequences, yielding 50{,}000 testing samples. Each testing sample serves as a source, with 10 reference samples randomly selected; the corresponding ground-truth outputs are available for direct comparison.

\noindent \textbf{Implementation Details}.
SynthCloner is implemented in PyTorch and trained on a single NVIDIA RTX 6000 Ada GPU for 400k steps. The model processes 1-second audio segments at 44.1 kHz using the AdamW optimizer \cite{adamw} with an initial learning rate of $10^{-4}$ and an exponential decay rate of $0.999996$. The batch size is 8, and the RVQ module uses a 1024-entry codebook with 8 quantization layers. We use same loss functions in FACodec \cite{facodec}. For reconstruction, 
we apply a multi-scale mel-spectrogram loss with 7 scales using FFT window lengths of [32, 64, 128, 256, 512, 1024, 2048]  paired with mel bins of [5, 10, 20, 40, 80, 160, 320], respectively. For each scale, the hop length is set to $\text{window\_length} / 4$. The total training objective is a weighted sum of 
multi-scale mel spectrogram loss $\lambda_{\text{mel}}\mathcal{L}_{\text{mel}}$, 
feature matching loss $\lambda_{\text{feat}}\mathcal{L}_{\text{feat}}$, 
adversarial loss $\lambda_{\text{adv}}\mathcal{L}_{\text{adv}}$, 
commitment loss \cite{van2017neural} $\lambda_{\text{commit}}\mathcal{L}_{\text{commit}}$, 
codebook loss \cite{van2017neural} $\lambda_{\text{codebook}}\mathcal{L}_{\text{codebook}}$, 
and classification losses for timbre, content, and ADSR envelope
($\lambda_{\text{timbre}}\mathcal{L}_{\text{timbre}} + \lambda_{\text{content}}\mathcal{L}_{\text{content}} + \lambda_{\text{adsr}}\mathcal{L}_{\text{adsr}}$).
We set $\lambda_{\text{mel}}=15.0$, $\lambda_{\text{feat}}=2.0$, 
$\lambda_{\text{adv}}=1.0$, $\lambda_{\text{commit}}=0.25$, 
$\lambda_{\text{codebook}}=1.0$, and 
$\lambda_{\text{timbre}}=\lambda_{\text{content}}=\lambda_{\text{adsr}}=5.0$.

\subsection{Evaluation Setting}
\noindent \textbf{Baseline Models}.
Since no prior work addresses SAT directly, we compare SynthCloner against two many-to-many timbre transfer models:  
(1) SS-VAE~\cite{ssvae}, a semi-supervised variational autoencoder for controllable audio generation, and  
(2) Control Transfer Diffusion (CTD)~\cite{cac}, a diffusion-based method for conditional timbre transfer. 
We further evaluate a variant of our model without the ADSR envelope path, keeping the ADSR envelope fixed during perturbations in other paths to assess the impact of explicit ADSR modeling.
Both the proposed and baseline models are trained on our SynthCAT dataset from scratch.

\noindent \textbf{Objective Metrics}.
We evaluate audio transfern with three metrics. Spectral fidelity is measured by the multi-scale STFT loss (MSTFT)~\cite{ddsp} from 
audiotools \cite{audiotools}
, which computes L1 distances over magnitude and log-magnitude spectra using STFT windows of 2048 and 512. ADSR envelope accuracy is assessed by the log-RMS distance (LRMSD), the L1 difference between log-RMS energy contours of the prediction and reference. Content accuracy is measured by the F0 root mean square error (F0RMSE), with pitch contours extracted via TorchCrepe \cite{torchcrepe}. 

\noindent \textbf{Subjective Metrics}.
We assess perceptual quality through Mean Opinion Score (MOS) tests on three dimensions: timbre similarity \textbf{(TMOS)} and ADSR envelope similarity \textbf{(ADSRMOS)}, measured between each prediction and its reference, and content similarity \textbf{(CMOS)}, measured between each prediction and its source. Listeners rated samples on a five-point Likert scale from 1 (``bad'') to 5 (``excellent''). Four source–reference pairs were randomly selected; for each pair, all models produced 4 predictions plus the ground truth (5 samples total). Twenty subjects rated all samples, and MOS with 95\% confidence intervals were reported.

\section{Experiment Results}
\subsection{Synthesizer-Style Audio Transfer}
As shown in Table 1, SynthCloner achieves the best performance across both objective and subjective metrics. The low MSTFT confirms that the predicted outputs closely match the ground-truth signals in spectral structure, demonstrating high fidelity in SAT. ADSR envelope accuracy follows the same trend: LRMSD remains low, consistent with faithful reproduction of target ADSR envelopes. Finally, content preservation is reflected by a low F0RMSE, suggesting that pitch trajectories are stable when timbre and ADSR are altered. Compared to the baselines, SS-VAE and CTD exhibit higher MSTFT and LRMSD and larger F0 deviations, pointing to pitch instability when ADSR envelope and timbre are manipulated jointly.
The subjective results corroborate the objective metrics. SynthCloner achieves the highest MOS across all dimensions, approaching ground truth, showing that listeners perceived its outputs as highly similar to the reference in timbre and ADSR envelope while preserving content. 

When trained without the ADSR envelope path, LRMSD rises notably and MSTFT increases, confirming that envelope dynamics cannot be captured without explicit modeling and indicating degraded spectral fidelity. However, F0RMSE remains relatively close to the full model, showing pitch content is largely unaffected. The subjective metrics reinforce this: ADSRMOS drops dramatically and TMOS decreases substantially without the envelope path, while CMOS shows a more modest decline, consistent with preserved pitch content. These findings underscore that explicit ADSR envelope modeling is crucial for effective SAT, significantly impacting spectral fidelity and perceptual quality while maintaining content integrity.


\subsection{Ablation Study}
In this section, we evaluate the independent attribute control of SynthCloner by 
three settings: (1) \textbf{Proposed}, which performs both timbre and ADSR envelope conversion, (2) \textbf{w/o timbre conv.}, which uses the input timbre instead of the reference, and (3) \textbf{w/o ADSR conv.}, which uses the input ADSR envelope instead of the reference. Relative to Proposed, \textbf{w/o timbre conv.} yields higher MSTFT, indicating that timbre conversion is crucial for spectral fidelity, while its LRMSD is only slightly worse, suggesting that envelope transfer remains unchanged. In contrast, \textbf{w/o ADSR conv.} shows much higher LRMSD than Proposed, demonstrating that envelope matching fails without explicit ADSR conversion.
For both partial conversions, F0RMSE is slightly worse than Proposed, indicating that pitch content is largely preserved but less stable when either attribute is missing.

\begin{table}[t!]
\centering
\setlength{\tabcolsep}{6pt}
\renewcommand{\arraystretch}{1.2}
\begin{tabular}{@{}lccc@{}}
\toprule
 & MSTFT $\downarrow$ & LRMSD $\downarrow$ & F0RMSE $\downarrow$ \\
\midrule
Proposed & \textbf{3.00} & \textbf{0.17} & \textbf{20.64} \\
w/o timbre conv.& 5.97 & 0.19 & 24.54 \\
w/o ADSR conv.& 4.15 & 0.39 & 24.06 \\
\bottomrule
\end{tabular}
\caption{ Objective result of independent attribute control.}

\label{tab:disent_comp}
\end{table}

\section{Conclusion}
\label{sec:concl}

We presented SynthCloner, the first model for synthesizer-style audio transfer that disentangles ADSR envelope, timbre, and content. We also introduced SynthCAT, a dataset with a rendering pipeline tailored for this task. Experiments show that SynthCloner outperforms baselines across all metrics, and ablations underscore the necessity of explicit ADSR modeling. Future work will aim to improve generalization to unseen timbres and extend the model to more complex configurations, including LFO modulations, additional envelope routing, and coupled effects.

\section{Acknowledgments}
This work was supported by the National Science and Technology Council (NSTC) of Taiwan under Grant NSTC 114-2628-E-002-013-MY3, and by the Ministry of Education of Taiwan. Funding to attend this conference was provided by the Carnegie Mellon University Graduate Student Assembly / Office of the Provost Conference Funding.




{\fontsize{9}{9.5}\selectfont
\bibliographystyle{IEEEbib}
\bibliography{strings,refs}
}
\end{document}